# Selective surface functionalization at regions of high local curvature in graphene

Qingzhi Wu,[a,b] Yaping Wu,[b] Yufeng Hao,[b] Jianxin Geng,[c] Matthew Charlton,[b] Shanshan Chen,[b] Yujie Ren,[b] Hengxing Ji,[b] Huifeng Li,[b] Danil W. Boukhvalov,[e] Richard D. Piner,[b] Christopher W. Bielawski,*[c,d] and Rodney S. Ruoff*[b]

**Monolayer graphene was deposited on a Si wafer substrate decorated with $SiO_2$ nanoparticles (NPs) and then exposed to aryl radicals that were generated *in situ* from their diazonium precursors. Using micro-Raman mapping, the aryl radicals were found to selectively react with the regions of graphene that covered the NPs. The enhanced chemical reactivity was attributed to the increased strain energy induced by the local mechanical deformation of the graphene.**

On account of its exceptionally high carrier mobility, thermal conductivity, and mechanical properties, graphene has attracted considerable attention for use in nanoelectronics, supercapacitors and thermal management applications, as well as in biological and chemical sensors, among others.[1-4] However, the zero bandgap of graphene is a drawback for applications that require controlled conductivities, such as logic gates. Thus, there is interest in the synthesis of quasi one-dimensional graphene nanoribbons, which confine the carriers to specified regions with defined band gaps.[5-8] While chemical doping provides a means to achieve such goals, the doping process introduces defects in a random fashion.[9-13] Indeed, the selective spatial functionalization of graphene remains an important challenge for the field.

It is generally accepted that the carbon atoms located within the plane of graphene are relatively chemically inert due to conjugation, while those located at the edges or at defects are more reactive.[14,15] However, according to 'π-orbital axis vector' (POAV) theory, the carbon atoms that reside on highly curved surfaces exhibit increased chemical potential due to diminished electronic delocalization.[16-18] They also have higher strain energy so that increased reactivity may be expected[19,20] and even described using the strain coordinates.[21,22] A simulation of the chemical reactivity of corrugated graphene by DFT predicted an enhancement if the ratio of height of the corrugation (ripple) to its radius was larger than 0.07.[23] Therefore, we envisioned that selectively deforming graphene would render the carbon atoms in the regions of high curvature chemically reactive.[24] To test this hypothesis, graphene was 'suspended' locally on $SiO_2$ nanoparticles to introduce local curvature and then exposed to aryl radicals to determine if such locally curved surfaces were more reactive than the planar regions.

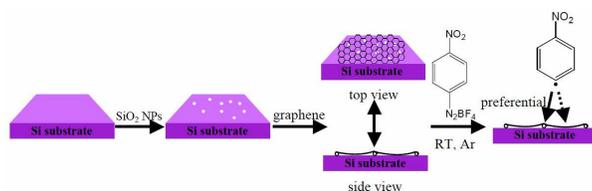

**Scheme 1** The functionalization of graphene with aryl radicals derived near $SiO_2$ nanoparticles decorating a Si substrate.

As illustrated in Scheme 1, CVD-grown graphene was transferred to a Si substrate having a 280-nm thick $SiO_2$ layer and then treated with 4-nitrophenyl diazonium tetrafluoroborate, which is known to decompose into aryl radicals. Fig. 1a and 1b show the Raman spectra of graphene on the Si substrate before and after treatment with the diazonium salt. The characteristic signals at ca. 1580 and 2670 $cm^{-1}$ were assigned to the G band and 2D bands which derive from the in-plane optical vibration and second-order zone boundary phonons, respectively. The signal at ca. 1350 $cm^{-1}$, which was attributed to the D band derived from first-order zone boundary phonons, was absent in the Raman spectra of graphene before chemical treatment but was clearly observed after treatment. The presence of the D band in the Raman spectra of graphene exposed to 4-nitrophenyl diazonium tetrafluoroborate has been previously shown to be due to the covalent attachment of nitrobenzene.[11,12,25,26] Moreover, the intensity of the D band was found to increase both over time and in the presence of higher concentrations of the diazonium salt, a result consistent with the transformation of $sp^2$ to $sp^3$ bonding of some of graphene's carbon atoms.[13,25-27] It is noteworthy that signal intensity in the region between the D band and G band also significantly increased after treatment, providing additional evidence for the change in hybridization.[11,27]

The covalent bonding of nitrobenzene with graphene was further assessed with XPS and the key results are summarized in Fig. 1c and 1d. The C1s signal observed at ca. 284.5 eV prior to exposure to the diazonium salt broadened and decreased in intensity after exposure, and was successfully fit by two Gaussian curves with maxima at ca. 284.5 and

285.5 eV (inset in Fig. 1c). The signal observed at ca. 285.5 eV was attributed to the C-N bond of the nitrobenzene group, while the decrease in the intensity of the peak at 284.5 eV was attributed to the decreased

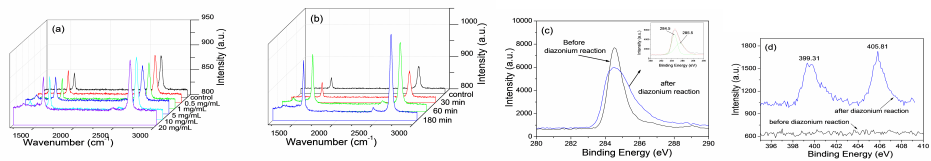

**Fig. 1** (a) Raman spectra of graphene before and after chemical treatment using various concentrations of the diazonium salt. (b) Raman spectra of graphene before and after the diazonium reaction, for different reaction times. (c) C1s XPS spectrum of the graphene before and after treatment (10 mg/mL of the diazonium salt in acetonitrile, 180 min). Inset shows the peak fitting of the C1s spectrum after the treatment. (d) N1s XPS spectrum of the graphene before and after treatment (10 mg/mL, 180 min).

number of $sp^2$ carbons in the lattice on account of the change in hybridization state.[28-30] The two signals observed at ca. 399.3 and 405.8 eV (Fig. 1d) were assigned to the aryl nitro group.[28-30] No significant signals from B or F were detected after chemical treatment (data not shown).

Next, the spatial distribution of nitrobenzene was assessed using two dimensional (2-D micro-Raman mapping in the D band region (1300-1400 cm$^{-1}$). The optical image of monolayer graphene on a Si substrate is shown in Fig. 2a. Several wrinkles were observed and are believed to form during the cool-down process on account of the different thermal expansion coefficients of graphene and Cu foil.[30] The Raman map marked with the frame in Fig. 2a generally showed low signal intensity except at some wrinkles (Fig. 2b), demonstrating that the CVD-grown graphene was of high quality and that the D defects were preferentially formed at these high curvature sites. Fig. 2d-f show three typical Raman spectra of points at the planar or wrinkled sites of graphene before and after chemical treatment. At the planar site (point 1), no obvious D band was detected before or after treatment (Fig. 2d). On the contrary, the intensity of the D band at the wrinkled sites (points 2 and 3) was significantly increased (i.e., a measured $I_D/I_G$ ratio of ca. 0.86 and 0.88) after treatment with the diazonium salt. The increase of the $I_D/I_G$ ratio along with the wrinkles after the diazonium reaction was also observed in the Raman map of the $I_D/I_G$ ratio (Fig. S1). Collectively, these results suggested to us that the reactivity of the carbon atoms along the wrinkles of graphene was higher than that in the relatively planar regions.

To further evaluate the curvature-induced enhancement of chemical reactivity of graphene, SiO$_2$ nanoparticles (NPs) with an average diameter of ca. 50 nm were deposited on the surface of a Si substrate before the graphene was transferred. As shown in Fig. 3, SEM revealed that the graphene featured wrinkles with different degrees of curvature as well as long range grain boundaries; aggregates of the SiO$_2$ NPs were also observed (as white dots). The variation in size of the wrinkles was presumed to be due to the propensity of the NPs to form clusters of various

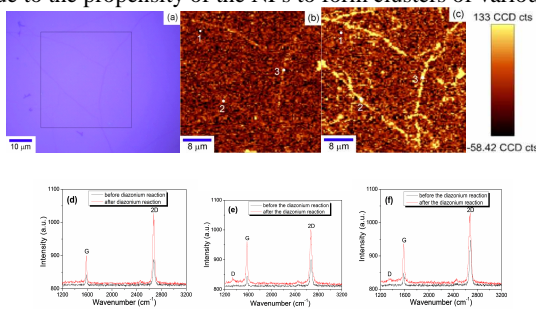

**Fig. 2** (a) Optical image of graphene on a Si substrate. (b) and (c) 2-D Raman map in the D band region (1300-1400 cm$^{-1}$) of graphene marked with the frame in before and after chemical treatment (10 mg/mL of the diazonium salt in acetonitrile, 60 min). (d)-(f) Raman spectra of graphene before and after chemical treatment, corresponding to points marked with 1, 2, and 3 in (b) and (c), respectively.

sizes. As shown in Fig. 4a, the Raman map of a SiO$_2$ NP perturbed graphene sheet exhibited an overall low D band intensity before treatment with the diazonium salt although the influence of the SiO$_2$ NPs was apparent (Fig. 4b). After treatment, a significant enhancement of the D band intensity was observed, particularly in the regions along the wrinkles. Fig. 4d and 4f show the Raman spectra of the points marked in Fig. 4b and 4c before and after exposure to the diazonium salt, respectively. At point 1, no significant increase of the D band intensity was observed after treatment, while a significant increase of the D band intensity was observed at points 2 and 3 (corresponding to $I_D/I_G$ ratios of ca. 0.85 and

0.86, respectively), indicating that the formation of more 'defects' at these curved sites was due to the covalent attachment of nitrobenzene. In support of this assessment, Raman mapping of the $I_D/I_G$ ratio showed an increase of the $I_D/I_G$ ratio along with the wrinkles after treatment with the diazonium salt (Fig. S2).

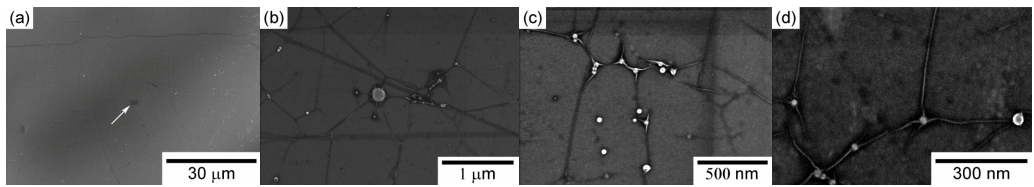

**Fig. 3** SEM images of graphene on a Si substrate decorated with $SiO_2$ nanoparticles. (b)-(d) are higher magnification images of the area marked with the white arrow in (a). Numerous wrinkles induced by the $SiO_2$ nanoparticles are visible. The shadow marked with the white arrow in (a) was attributed to the extensive irradiation by the electron beam during imaging.

It is known that carbon atoms at the edges and defects of graphene are more reactive than those in the basal plane.[14,15] However, as we have previously outlined[19,20] and showed above,

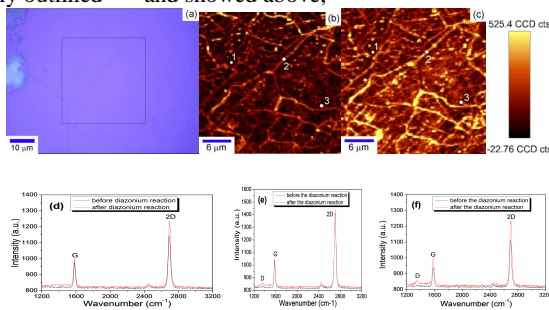

**Fig. 4** (a) Optical image of graphene on a Si substrate decorated with $SiO_2$ nanoparticles. (b) and (c) 2-D Raman maps of the D-band intensity of graphene marked with the frame in before and after chemical treatment (10 mg/mL diazonium salt in acetontrile, 60 min). (d)-(f) Typical Raman spectra of graphene before and after treatment, corresponding to the points marked with 1, 2, and 3 in (b) and (c), respectively.

the carbon atoms in highly curved (i.e., highly deformed) regions are more reactive. Theoretical predictions and experimental observations have demonstrated that the electronic properties of graphene may change through the application mechanical strain.[31-35] For example, under uniaxial tensile stress, a theoretical band-gap of ca. 300 meV for graphene was calculated, with significant red shifts of the G and 2D bands observed in its Raman spectrum.[31] Likewise, when grown on a Pt (111) surface, graphene nanobubbles formed which gave rise to local 'pseudo-magnetic fields' greater than 300 T.[34] To build on these results and to help elucidate the mechanism of the radical coupling reaction observed above, the covalent binding of nitrophenyl groups to flat and rippled graphene surfaces were calculated using DFT; in parallel, the pyramidalization angles of the respective ripples were measured via POAV theory. The calculations demonstrated that increasing the curvature of the ripples resulted in increased pyramidalization that was concomitant with decreased nitrophenyl binding energy (see Fig. S3). Moreover, the calculations showed that increased surface reactivity was observed when the corrugation height to radius ratio was greater than 0.15, in accord with previous data.[23] Therefore, although the mechanism of the aryl radical addition remains under investigation,[36] it appears from our work that the reaction may be guided through the development of local mechanical deformation and other intimately coupled phenomena, including changes in the density of states and electronic chemical potential.

In summary, transferring graphene onto a Si wafer substrate decorated with $SiO_2$ NPs induced local regions of mechanical strain and increased the chemical reactivity of at least some of the carbon atoms at these sites. In particular, *in situ* generated aryl radicals were found to couple to the graphene with a higher degree of local curvature, as confirmed by micro-Raman mapping spectroscopy. It is suggested that future studies use substrates that will introduce deformation in well defined areas (e.g., patterned substrates with high aspect features) to not only enable the chemical patterning of graphene but also to enhance its chemical reactivity in a selective manner.

We appreciate advice from K. Ausman. We acknowledge support from the NSF (DMR-0907324), the CSC, the National Natural Science Foundation of China (Grant No. 30800256), and the Program for Changjiang Scholars and Innovative Research Team (IRT1169) at the Wuhan University of Technology. CWB acknowledges the Welch Foundation (F-1621), the ARO (W911NF-09-1-0446), and the WCU program (R31-10013) administered through the

NRF of Korea and funded by the Ministry of Education, Science, and Technology. DWB acknowledges computational support from the CAC of KIAS.

## Notes and references

*a* State Key Laboratory of Advanced Technology for Materials Synthesis and Processing, and Biomedical Materials and Engineering Center, Wuhan University of Technology, Wuhan 430070, China

*b* Department of Mechanical Engineering and the Materials Science and Engineering Program, The University of Texas at Austin, 1 University Station C2200, Austin, TX, 78712, USA. E-mail: r.ruoff@mail.utexas.edu.

*c* Department of Chemistry and Biochemistry, The University of Texas at Austin, 1 University Station A1590, Austin, TX 78712, USA. E-mail: bielawski@cm.utexas.edu

*d* World Class University (WCU) Program of Chemical Convergence for Energy & Environment (C2E2), School of Chemical and Biological Engineering, Seoul National University, Seoul 151-742, Korea

*e* School of Computational Sciences, Korea Institute for Advanced Study, Seoul 130-722, Korea.

† Electronic Supplementary Information (ESI) available: Additional experimental details, density function theory (DFT) calculations, and Raman spectra. See DOI: 10.1039/b000000x/

# Selective surface functionalization at regions of high local curvature in graphene

*Qingzhi Wu, Yaping Wu, Yufeng Hao, Jianxin Geng, Matthew Charlton, Shanshan Chen, Yujie Ren, Hengxing Ji, Huifeng Li, Danil W. Boukhvalov, Richard D. Piner, Christopher W. Bielawski,\* and Rodney S. Ruoff\**

**Synthesis and Transfer of Graphene.** Large-area, high-quality graphene monolayers were synthesized using the CVD method reported previously (*Science,* 2009, **324**, 1312) and transferred to different substrates (*Nano Lett.,* 2009, **9**, 4359). In a typical process, a solution of PMMA in chlorobenzene (350 kDa, 46 mg/mL) was spin-coated on the surface of graphene. Then, after the Cu substrate was completely dissolved in $(NH_4)_2S_2O_8$ aqueous solution (0.5 M), the PMMA-graphene material was transferred to the surface of a Si substrate (280-nm thick $SiO_2$ layer) or onto the same type of Si substrate decorated with $SiO_2$ nanoparticles of ~50 nm diameter. The PMMA-graphene-Si samples were dried in air for 30 min and then under vacuum ($10^{-2}$ Torr) for 30 min to remove residual water. Finally, the PMMA was removed for each type of sample using acetone.

**Chemical Treatment of Graphene.** In a typical process, graphene on the Si wafer substrate was immersed into an acetonitrile solution of 4-nitrophenyl diazonium tetrafluoroborate at known concentrations and for known periods of times. The chemical reaction between graphene and this diazonium salt was carried out at room temperature in the dark and in a glove box (Ar atmosphere). Afterward, the sample was washed with acetonitrile several times and dried.

**Microscopy and Spectroscopy.** All substrates supported with graphene were characterized via optical microscopy (Zeiss Axioskop), scanning-transmission electron microscopy (Hitachi S-5500), and micro-Raman spectroscopy (WITEC Alpha 300, =488 nm, 100× objective lens). The elemental analysis (C, N, B, F) and chemical bonding analysis of graphene was conducted using a Kratos AXIS Ultra DLD X-ray photoelectron spectrometer.

**Density Functional Theory (DFT) Calculations.** The calculations were performed for large supercells (128 carbon atoms) that provided a proper separation of the ripples. All technical details of the calculations may be found in the following reference: *J. Phys. Chem. C* 2009, **113**, 14176-14178. The pyramidization angle calculations were performed using the method described in *J. Phys. Chem. A* 2001, **105**, 4164-4165. The calculations demonstrate that increasing the pyramidization angle (i.e., increasing the curvature of the ripples) increases binding energies and decreases the covalent bond formation energy (Fig. S3b). The changes in the total energy of graphene under the formation of small ripples were comparable with the strain energy above 5% (see *ChemPhysChem* 2012, **13**, 1463–1469). The value of the binding energies for a pair of nitrophenyl groups to flat graphene was about 2.5 eV, which is comparable to that calculated for hydrogen (to form graphane from graphene). Increasing the pyramidization angles lead to increased binding energies such that the values were comparable to the calculated values of the vacancy formation in flat graphene (about 7 eV). Estimated temperature required for the desorption of nitrophenyl groups from the top of the ripples was > 1000 K.

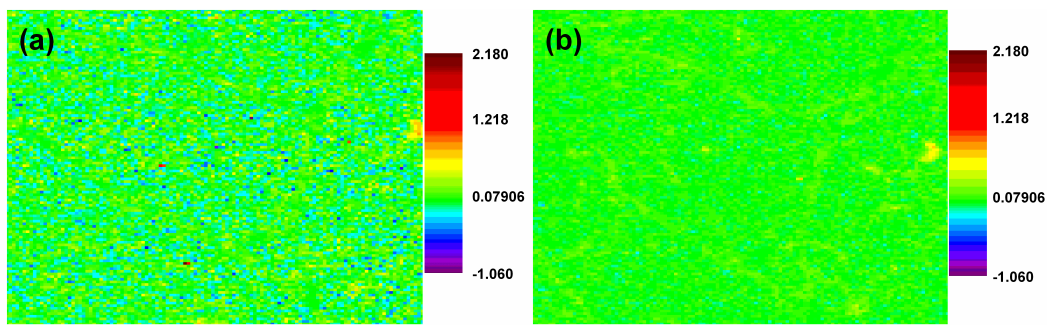

**Figure S1.** Raman maps of the D/G ratio of graphene on a Si substrate (a) before and (b) after treatment with 10 mg/mL of the diazonium salt in acetonitrile for 60 min.

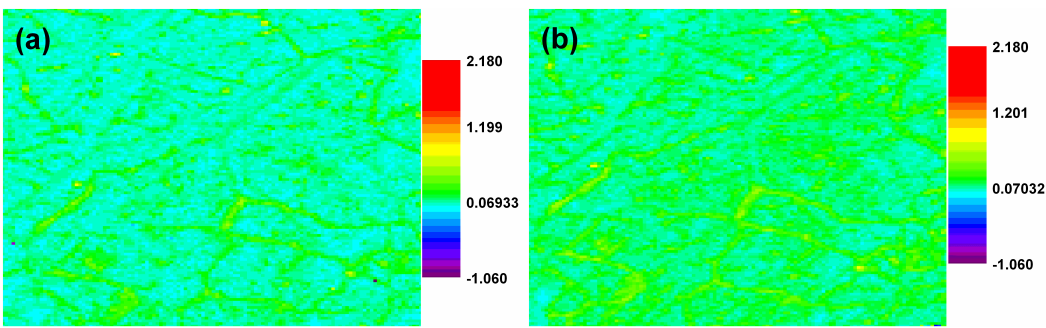

**Figure S2.** Raman maps of the D/G ratio of graphene on a Si substrate decorated with $SiO_2$ nanoparticles (a) before and (b) after treatment with 10 mg/mL of the diazonium salt in acetonitrile for 60 min.

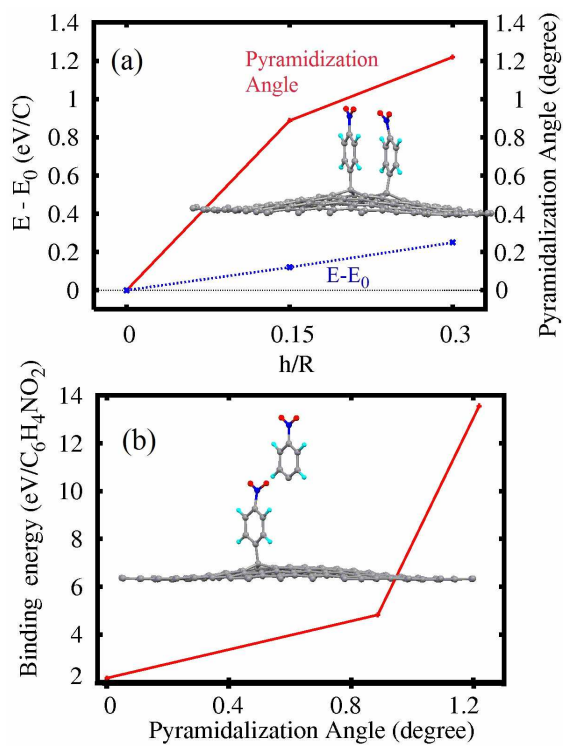

**Figure S3.** (a) Calculated pyramidization angles (solid red line) and total energy per carbon atom of pure graphene as function of the height (h)/radii (R) ratio of the ripples. (b) Binding energy for a pair of p-nitrophenyl groups as a function of the pyramidization angle. The bond formation energy of the p-nitrophenyl groups to flat graphene was set to zero. The structture shown in the inset reflects the optimized atomic structures of the p-nitrophenyl groups on the ripple with h/R = 0.15.